\documentclass[letterpaper]{article} 
\usepackage{aaai25}  
\usepackage{times}  
\usepackage{helvet}  
\usepackage{courier}  
\usepackage[hyphens]{url}  
\usepackage{graphicx} 
\usepackage{natbib}  
\usepackage{caption} 
\usepackage[ruled,vlined]{algorithm2e}

\usepackage{amsmath,amsfonts,amssymb,amsthm}
\usepackage{subfig}
\usepackage{bm}
\usepackage{enumitem}
\usepackage{multirow}
\usepackage{booktabs}
\usepackage{mathtools}
\usepackage{color}

\newcommand{\ie}{\textit{i.e.}}
\newcommand{\eg}{\textit{e.g.}}

\newcommand{\adj}{\tilde{\mathbf{A}}}
\newcommand{\emb}{\mathbf{E}}

\newcommand{\concat}{\mathop{\big{\|}} \limits}
\newcommand{\mlu}[1]{{#1}^{(m)}}
\newcommand{\ulu}[1]{{#1}^{(u)}}
\newcommand{\ilu}[1]{{#1}^{(i)}}

\title{
STAIR: Manipulating Collaborative and Multimodal Information for E-Commerce Recommendation
}
\author{
    Cong Xu\equalcontrib,
    Yunhang He\equalcontrib,
    Jun Wang\footnotemark[2],
    Wei Zhang\thanks{Corresponding authors.}
}
\affiliations{
    East China Normal University \\

    Shanghai, China\\
    \{congxueric, yhhe2004, wongjun, zhangwei.thu2011\}@gmail.com
}

\usepackage{bibentry}

\begin{document}

\maketitle

\begin{abstract}
While the mining of modalities is the focus of most multimodal recommendation methods, 
we believe that how to fully utilize both collaborative and multimodal information is pivotal in e-commerce scenarios
where, as clarified in this work, the user behaviors are rarely determined entirely by multimodal features.
In order to combine the two distinct types of information,
some additional challenges are encountered:
1) Modality erasure: 
Vanilla graph convolution, which proves rather useful in collaborative filtering, however erases multimodal information;
2) Modality forgetting: Multimodal information tends to be gradually forgotten as the recommendation loss essentially facilitates the learning of collaborative information.
To this end,
we propose a novel approach named STAIR,
which employs a novel \underline{ST}epwise gr\underline{A}ph convolution to enable a co-existence of collaborative and multimodal \underline{I}nformation in e-commerce \underline{R}ecommendation.
Besides, it starts with the raw multimodal features as an initialization,
and the forgetting problem can be significantly alleviated through constrained embedding updates.
As a result,
STAIR achieves state-of-the-art recommendation performance on three public e-commerce datasets 
with minimal computational and memory costs.
Our code is available at \url{https://github.com/yhhe2004/STAIR}.
\end{abstract}

\section{Introduction}

Recommender systems are playing an increasingly important role nowadays \cite{chen2019behavior,el2022twhin}
as the explosion of data has far exceeded the human's retrieval capabilities.
Unlike traditional data mining tasks \cite{hoffmann2007kernel}, 
recommender systems are rapidly growing to adapt to different interaction forms and data types.
Consequently multimodal recommendation research~\cite{he2015vbpr,wei2023mmssl} is becoming a popular topic since textual and visual modalities are the main elements that appear in modern media.
Nonetheless, their utility usually depends on the relevant recommendation scenario~\cite{mcelfresh2022generalizability}.
For example, textual features in news recommendation and visual features in video recommendation can be highly instrumental 
because they are indeed the core that attracts users.
However, this is often not the case for non-content-driven recommendation scenarios such as e-commerce,
where user interaction behaviors may be influenced by a variety of factors beyond modalities.

While numerous approaches~\cite{wei2019mmgcn,Zhou2023bm3,wei2023mmssl} have been proposed to mine the multimodal features,
the true utility of raw multimodal features in e-commerce datasets has not yet been clarified.
We are to ascertain in this scenario whether the user behaviors are driven by a certain type of modality.
By clustering features to infer the category of each item, 
the modal-specific user behavior uncertainty can be estimated via Shannon entropy.
The average uncertainty across all users then can be used as an indicator.
Unfortunately, we find that the behavior uncertainty under textual or visual modalities is much worse than that under pre-trained embeddings; 
sometimes even only slightly better than randomly generated Gaussian noise.
This suggests that interaction behaviors in these e-commerce datasets are often not driven by textual or visual content. 
Hence these multimodal features should be used as a complement to, rather than a substitute for, collaborative information.
\textit{For this reason, 
we believe that how to fully utilize both collaborative and multimodal information becomes pivotal.}
Unlike previous work that learns embeddings from scratch~\cite{zhang2021lattice,zhou2023freedom}, 
STAIR starts with multimodal features as an initialization and seeks to progressively combine these two distinct types of information together through some special designs.
In practice, we have encountered two additional challenges.

Graph convolutional networks (GCNs)~\cite{kipf2016gcn,he2020lightgcn} have become the de facto architecture in collaborative filtering,
among which LightGCN~\cite{he2020lightgcn} is a pioneering work known for its simplicity and effectiveness.
When these advanced techniques are applied to the modality-initialized embeddings, we, however, observe a severe \textit{modality erasure} phenomenon:
as illustrated in Figure 3, the increase in GCN layers leads to the incorporation of more collaborative information; however, this simultaneously results in the rapid erosion of modality information.
In particular,
unlike the over-smoothing issue~\cite{li2018oversmoothing} that often occurs after stacking multiple convolutional layers,
this erasure phenomenon can be observed as soon as the embedding has been convolved.
To balance the trade-off between collaborative and multimodal information, we develop a stepwise convolution mechanism.
Note that the generalized LightGCN over the entire embeddings
is equivalent to convolving each dimension independently and then concatenating them together.
By assigning a varying layer weight along the embedding dimension, 
the preceding dimensions will contain enough collaborative information, 
and conversely, the later dimensions will retain more multimodal information.

Both types of information can now co-exist within the embeddings through the proposed stepwise graph convolution.
However, it is still inevitable that the multimodal features (incorporated via initialization) will be forgotten during the training process,
since the training objective (\eg, BPR loss~\cite{rendle2012bpr}) is designed to encourage the interacted pairs to be closer in the embedding space,
which is actually to promote collaborative information learning.
Statistically, we observe from Figure 5 that the Pearson correlation between the embeddings and the modality initialization declines rapidly.
Inspired by recent advancements in constrained embedding learning~\cite{xu2024sevo},
we enforce the most modality-similar items to be updated in a consistent manner.
This actually performs a backward stepwise convolution over the descent directions, 
in which the layer weights vary along the embedding dimension in a manner opposite to the forward process.
As a result, the later dimensions will receive greater modality enhancement compared to the preceding ones.

The main contributions of this paper can be summarized as follows.
\begin{itemize}[leftmargin=*]
  \item To the best of our knowledge, we are the first to identify the poor modal-specific user behavior uncertainty in e-commerce scenarios.
  We also empirically validate that the vanilla graph convolution proves useful in traditional collaborative filtering however tends to erase multimodal information,
  leading to sub-optimal results.
  \item 
  We propose STAIR with modality initialization and forward stepwise graph convolution to address the aforementioned problems.
  Furthermore, the backward stepwise convolution is employed to mitigate the forgetting issue.
  \item
  STAIR achieves state-of-the-art recommendation performance over three public e-commerce datasets, with improvements ranging from 2\% to 6\%.
  More importantly, these gains can be achieved at only $1/2$ or even $1/100$ of the costs of other multimodal methods.
\end{itemize}

\section{Related Work}

\paragraph*{Traditional collaborative filtering} for personalized recommendation
primarily retrieves the items based on a user's historical interactions (\eg, clicking and purchasing behaviors).
The most classic collaborative filtering approach is the matrix factorization technique~\cite{rendle2012bpr,koren2008svd++},
which aims to embed users and items separately into a unified latent space for the subsequent score calculations. 
With the development of deep learning,
NCF~\cite{he2017ncf} and ENMF~\cite{chen2020emnf} introduce additional trainable modules to further enhance interaction modeling.
Note that the interaction data can be exactly represented as a bipartite graph, 
hence the growing research interest in graph convolutional networks (GCNs)~\cite{berg2017graph,wang2019ngcf,yang2022hicf}.
LightGCN~\cite{he2020lightgcn} is a pioneering work known for its simplicity and effectiveness.
Several recent studies~\cite{wu2019simplifying,shen2021smoothing} have clarified its superiority in incuring embedding smoothness,
with some work~\cite{guo2023jgcf} to improve this approach toward manipulating both low-frequency and high-frequency signals.
However, previous efforts focusing only on the interaction data struggle to accurately model embeddings for those inactive entities.
While the use of multimodal features is helpful in solving this problem, 
we find that existing GCNs are not sufficient to handle these multimodal features effectively.

\paragraph*{Multimodal recommendation}
have emerged as a significant area of research~\cite{he2015vbpr,wang2021dualgnn,wei2023lightgt}
with the surge of various modalities (\eg, text and image) in media.
In scenarios where user behaviors are significantly influenced by modalities (\eg, in news and video recommendations), 
the raw multimodal features can be readily utilized to complement collaborative signals.
MMGCN~\cite{wei2019mmgcn} employs a separate GCN for each modality to boost the personalized recommendation of micro-video.
In conjunction with cross-modal contrastive learning, MMSSL~\cite{wei2023mmssl} could achieve the state-of-the-art performance.
Nevertheless, the raw features appear to be less beneficial in non-content-driven scenarios like e-commerce.
Although some approaches, including LATTICE~\cite{zhang2021lattice} and the follow-up works~\cite{Zhou2023bm3,zhou2023freedom,yu2023mgcn,zhu2024modeling},
have somewhat recognized this problem and proposed some necessary modifications,
we \textit{quantitatively} justify for the first time the modal-specific user behavior uncertainty.
Furthermore, we also demonstrate that the vanilla graph convolution~\cite{he2020lightgcn} incuring smoothness 
have challenges in effectively integrating collaborative and multimodal information, leading to sub-optimal results.

\section{Methodology}

\begin{figure}[t]
    \centering
    \includegraphics[width=0.47\textwidth]{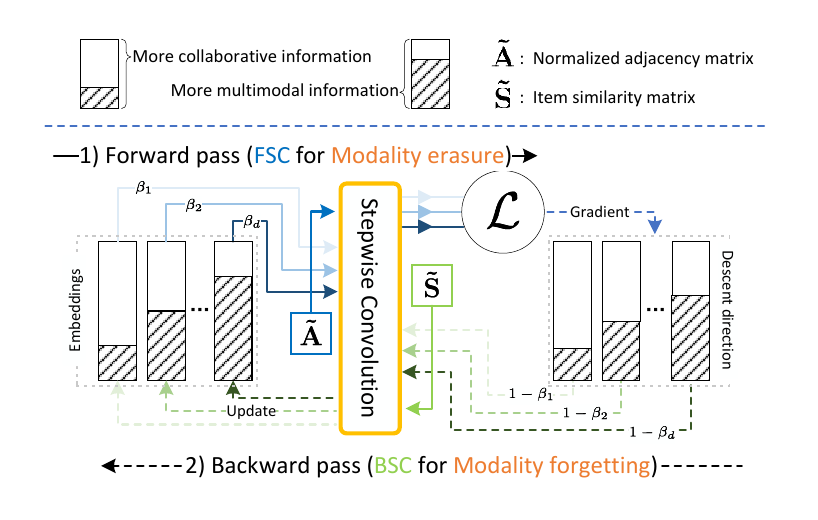}
    \caption{
      Overview of STAIR.
      1) \textbf{Forward pass}: More multimodal information are retained in the later dimensions via FSC.
      2) \textbf{Backward pass}: BSC is used to enhance multimodal information to mitigate modality forgetting problem.
    }
  \label{fig-framework}
\end{figure}

In this section, we first introduce the recommendation tasks we are interested in, and the motivation from our experimental observations.
Subsequently the challenges and corresponding modifications for the co-existence of collaborative and multimodal information will be presented in detail.
Figure~\ref{fig-framework} shows an overview of STAIR.

\paragraph*{Problem definition.}
Collaborative filtering~\cite{rendle2012bpr,he2020lightgcn} typically predicts an item $i \in \mathcal{I}$ 
that the user $u \in \mathcal{U}$ is most likely to prefer based on historical interactions.
To quantify how much a user likes an item,
they are first embedded into $d$-dimensional vector representations,
followed by the inner product usually serving as the score.
The top-$N$ candidate items then can be recommended to the user by ranking all items in descending order based on the scores.

Apart from interaction data, other side information such as multimodal features can be utilized to more adequately model item representations.
They complement the lack of objective characterization inherent in collaborative signals.
For instance, we can infer that two items are similar if they belong to the same brand (from text descriptions) or share the same style (from thumbnails).
The emergence of multimodal recommendation is to make full use of both collaborative and multimodal information.
Following previous works~\cite{zhang2021lattice,zhou2023freedom}, 
we focus on textual and visual modalities $\mathcal{M} = \{t, v\}$ with their encoded features $\mlu{\mathbf{F}} \in \mathbb{R}^{|\mathcal{I}| \times d_m}$, $m\in\mathcal{M}$.
Besides, let $\ulu{\emb} \in \mathbb{R}^{|\mathcal{U}| \times d}, \ilu{\emb} \in \mathbb{R}^{|\mathcal{I}| \times d}$ respectively represent the user/item ID embeddings,
and $\emb \in \mathbb{R}^{(|\mathcal{U}| + |\mathcal{I}|) \times d}$ the concatenation of $\ulu{\emb}$ and $\ilu{\emb}$.

\subsection{Motivation: Modal-Specific Behavior Uncertainty}

Before the multimodal features being extensively utilized for recommender systems~\cite{zhang2021lattice,zhou2023freedom,wei2023mmssl},
it is necessary to thoroughly investigate the true utility of these features.
Otherwise, these features may adversely affect recommendation performance particularly for e-commerce scenarios.

\begin{figure}[t]
    \centering
    \includegraphics[width=0.47\textwidth]{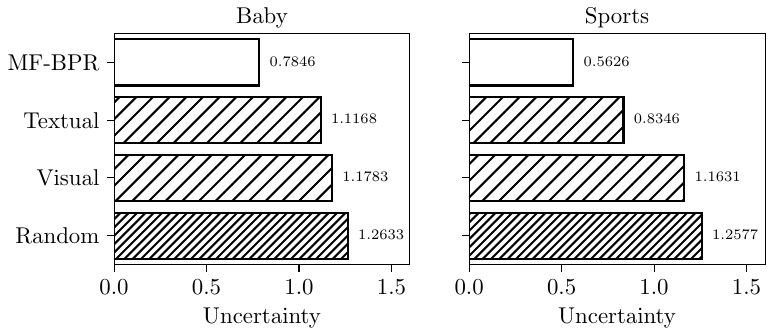}
    \caption{
        User behavior uncertainty under various features. 
        \textbf{MF-BPR}: the embeddings pre-trained via MF-BPR;
        \textbf{Textual/Visual}: the raw multimodal features;
        \textbf{Random}: the randomly generated Gaussian noise.
    }
  \label{fig-uncertainty}
\end{figure}

\paragraph*{High user behavior uncertainty in e-commerce.}
Multimodal features would be instrumental for personalized recommendations if they are dominant factors appealing to users.
However, users' purchasing decisions in e-commerce scenarios might be influenced by a variety of factors, not just modality-oriented.
If this is true, the multimodal features should be used as a complement to, rather than a substitute for, collaborative information. 
To figure this out, we employ an uncertainty metric~\cite{wang2024uncertainty} to quantify how inconsistent the user behaviors are in commonly used e-commerce datasets.
Specifically,
we perform K-Means~\cite{arthur2006kmeans} clustering
on each modality $\mlu{\mathbf{F}}, m \in \mathcal{M}$ separately, assigning a unique class label to each item. 
Then the Shannon entropy is used to compute the uncertainty surrounding a user. 
Intuitively, the behaviors for a user tend to exhibit low modal-specific uncertainty
if most of the corresponding interacted items fall into the same category.
The average behavior uncertainty across all users then can be used to depict the utility of the modality for this scenario.

Unfortunately, we find the multimodal features are not as attractive in e-commerce datasets (\eg, the widely used Baby and Sports).
We illustrate in Figure~\ref{fig-uncertainty} the respective user behavior uncertainty under textual or visual modalities.
For comparison, the embeddings pretrained via MF-BPR and random noise sampled from a standard Gaussian distribution are also included.
It makes sense that the pre-trained embeddings have the lowest level of uncertainty 
because the BPR loss encourages user embeddings to align with their neighboring item embeddings.
However, the multimodal features, particularly the visual modality, carry behavior uncertainty as poor as that under the random noise.
This verifies our assumption about e-commerce scenarios.

Since the user behaviors are rarely driven entirely by modalities, 
the focus of multimodal recommendation in e-commerce should be on how to fully utilize both collaborative and multimodal information.
In this work, we propose a novel approach named STAIR for this purpose.
Note that item embeddings in collaborative filtering are often trained from scratch.
Despite the inherent impurity of multimodal features, 
they can still offer a satisfactory initialization superior to random noise.
Hence, STAIR starts with multimodal features as an initialization and seeks
to progressively combine these two distinct types of information together through some special designs.
The first problem is how to compress these dimensionally inconsistent multimodal features into the same space without a well-trained projector. 
We turn to the whitening technique as proposed in \cite{su2021whitening,huang2021whitening}, 
which performs Singular Value Decomposition (SVD) on centralized features and retains the top-$d$ left singular vectors for item embedding initialization.
As for the user embedding, it is initialized by mean pooling its neighboring items.
Certainly, this straightforward initialization fails to preserve all valuable multimodal information. 
Subsequent modifications pose additional challenges, which we will introduce next.

\subsection{Modality Erasure and Graph Convolution}

\begin{figure}[t]
    \centering
    \includegraphics[width=0.45\textwidth]{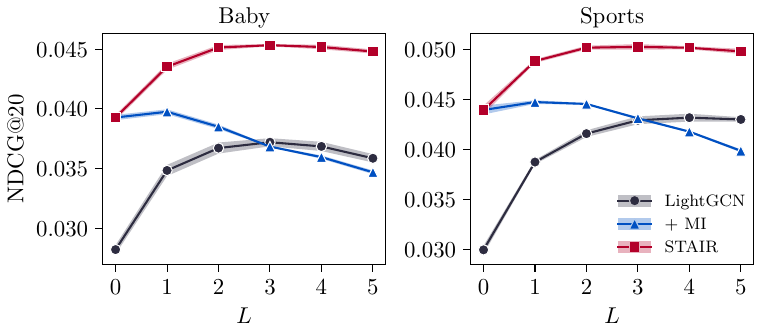}
    \caption{
        Modality erasure problem.
        Multimodal information (carried by Modality Initialization (+MI)) will be rapidly erased after a few convolution layers. 
    }
  \label{fig-erasure}
\end{figure}

\paragraph*{Challenge I: Modality erasure.}
Graph Convolutional Networks (GCNs) have become the de facto architecture (\eg, LightGCN) in the field of collaborative filtering 
due to their superiority in feature smoothing~\cite{shen2021smoothing}.
A $L$-layer LightGCN with equal layer weights can be formulate as follows
\begin{equation*}
    \mathbf{H} = \sum_{l=0}^L \frac{1}{L+1} \adj^l \emb,
\end{equation*}
where $\mathbf{A}$ denotes the adjacency matrix and $\adj$ is the corresponding matrix after symmetric sqrt normalization.
Unfortunately, this technique cannot be directly applied to $\emb$ which has incorporated multimodal information. 
Due to the smoothing nature of LightGCN, 
the valuable modality information will be rapidly erased as $L$ increases. 
This can be empirically verified from Figure~\ref{fig-erasure}. 
In contrast to the over-smoothing problem~\cite{li2018oversmoothing} that typically arises when $L > 3$, 
the modality erasure issue promptly occurs.
While both multimodal information from initialization and collaborative information from graph convolution are useful in their own ways, 
the challenge is how to effectively utilize both types of information, as they seem to conflict with each other.
To the best of our knowledge, nobody has explicitly pointed this out, 
but some approaches such as LATTICE~\cite{zhang2021lattice} and FREEDOM~\cite{zhou2023freedom} 
inadvertently choose to utilize two separate branches to bypass this problem.
In contrast, we develop a soft transition to effortlessly fuse both.

\paragraph*{Forward Stepwise Convolution (FSC).}
In order to handle the aforementioned trade-off, we develop a stepwise graph convolution mechanism.
This modification depends on the dimension-independent nature of the generalized LightGCN:
\begin{equation}
    \label{eq-fsc}
    \mathbf{H} =  \concat_{j=1}^d \mathbf{H}_{:, j}, \: \mathbf{H}_{:, j} = \sum_{l=0}^L \alpha_{jl} \adj^l \emb_{:, j},
\end{equation}
where $\emb_{:, j}$ stands for the $j$-th dimension of $\emb$ and $\concat$ is the concatenation operation.
Moreover, in contrast to the shared layer weights, 
the stepwise graph convolution assigns each dimension $j = 1,\ldots, d$ at layer $l = 0,\ldots, L$ a distinct weight:
\begin{equation} \label{eq_gamma}
    \alpha_{jl} = \frac{1 - \beta_j}{1 - \beta_j^{L+1}} \beta_j^l, \:
    \beta_j = 0.9 \cdot \bigg[1 - \Big(\frac{j-1}{d} \Big)^\gamma \bigg].
\end{equation}
Here, $\gamma$ is the sole hyperparameter that adjusts the weights of each dimension layer.
There are three noteworthy designs:
\textbf{1)} The value of $\alpha_{jl}$ is determined by the \textit{teleport ratio} $\beta_j \in [0, 1)$. 
Notice that the stepwise convolution returns the embedding itself if $\beta_j = 0$, 
and degenerates to a LightGCN as $\beta_j \rightarrow 1$.
Note that the formulation of $\beta_j$ w.r.t $j$ is not unique and other \textit{monotone} functions can be used.
The power function is chosen here because of its ease of tuning.
\textbf{2)} The layer weight $\alpha_{jl}$ diminishes as the layer $l$ increases.
A similar mechanism has been proposed in PageRank-inspired GCNs~\cite{gasteiger2018appnp,huang2021cs} to mitigate over-smoothing.
\textbf{3)} The normalization coefficient $(1 - \beta_j) / (1 - \beta_{j}^{L+1})$ is explicitly integrated to ensure that $\sum_{l=0}^L \alpha_{jl} = 1$ holds true. 
Consequently the magnitude across different dimensions remains at a consistent level.

\begin{figure}[t]
    \centering
    \includegraphics[width=0.45\textwidth]{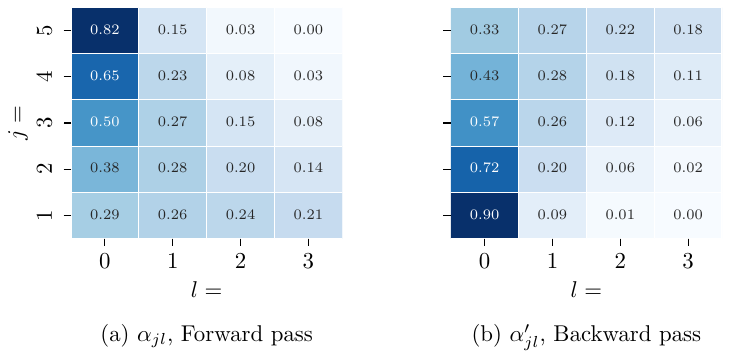}
    \caption{
        Stepwise layer weights under $\gamma=1$, $d=5$, and $L=3$.
        (a) $\alpha_{jl}$ for forward pass; (b) $\alpha_{jl}'$ for backward pass.
    }
  \label{fig-layer-weights}
\end{figure}

Figure~\ref{fig-layer-weights}a illustrates a toy example.
For the preceding dimensions (\eg, $j=1,2$), the teleport ratio $\beta_j$ is large and thus the layer weight is evenly distributed across each layer.
The stepwise convolution for these dimensions performs similarly to the traditional LightGCN. 
In contrast, multimodal information can be adequately preserved for the following dimensions (\eg, $j=4, 5$),
as the majority of layer weights are concentrated in the earlier convolutional layers.
\textit{
In a nutshell, 
by adjusting the hyperparameter $\gamma$ to a smaller (or larger) value, it is possible to contain more multimodal (or collaborative) information.
}

\subsection{Modality Forgetting and Enhancement}

\begin{figure}[t]
    \centering
    \includegraphics[width=0.43\textwidth]{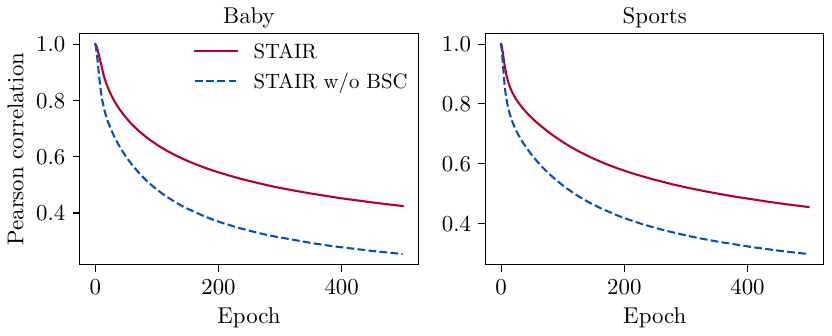}
    \caption{
        Modality forgetting problem.
        Pearson correlation coefficients between embeddings and multimodal features are averaged across all dimensions.
    }
  \label{fig-forgetting}
\end{figure}

\paragraph*{Challenge II: Modality forgetting.}
It is na\"ive to expect that the modality initialization only could capture all useful information inherent in multimodal features.
Note that the item embeddings $\ilu{\emb}$ is updated at step $t$ as follows:
\begin{equation*}
    \ilu{\emb}_{t} \leftarrow \ilu{\emb}_{t-1} - \eta \cdot \Delta \ilu{\emb}_t,
\end{equation*}
where $\Delta \ilu{\emb}_t$ denotes the descent direction and $\eta$ is the step size (\ie, learning rate).
For a recommendation training objective (\eg, BPR loss), 
$\Delta \ilu{\emb}_t$ directs the optimization to a region that promotes item embeddings closer to their respective neighboring user embeddings.
On the contrary, there is no mechanism to ensure that the initialized multimodal information can be consistently preserved,
leading to a potential modality forgetting issue~\cite{hu2024forgetting}.
This can be empirically validated through the Pearson correlation between trained embeddings and the initial multimodal features. 
As illustrated in Figure~\ref{fig-forgetting}, 
the original STAIR suffers from the forgetting problem during training, and the modification introduced next can significantly mitigate this.

\paragraph*{Backward Stepwise Convolution (BSC).}
Inspired by the recent advancements in constrained embedding learning~\cite{xu2024sevo}, 
the original descent direction $\Delta \ilu{\emb}_t$ is transformed so that the embeddings of items with high modality similarity are updated in a consistent manner.
Following \cite{zhang2021lattice,zhou2023freedom},
the modality similarity between two items is also depicted via the kNN graph extracted from multimodal features.
Recall from the Figure~\ref{fig-uncertainty} that
the various multimodal features typically suggest differing user behavior uncertainty.
It should identify more neighbors primarily from modalities of higher utility,
rather than selecting the same number of neighbors for all modalities as in previous work.
Formally, the $k_m$ neighbors for item $i$ of the modality $m \in \mathcal{M}$ are defined as follows
\begin{equation} \label{eq_knnk}
    \mathcal{N}_m(i) := \text{top-}k_m 
    \bigg(
    \bigg\{\frac{(\mlu{\bm{f}}_i)^T \mlu{\bm{f}}_j}{\|\mlu{\bm{f}}_i\| \| \mlu{\bm{f}}_j\| } \bigg\}_{j=1}^{|\mathcal{I}|}
    \bigg),
\end{equation}
where $\mlu{\bm{f}}_i \in \mathbb{R}^{d_m}$ denotes the $i$-th row vector of $\mlu{\mathbf{F}}$.
Subsequently, the overall similarity matrix $\mathbf{S} \in \mathbb{R}^{|\mathcal{I}| \times |\mathcal{I}|}$ is constructed 
by aggregating the information from the different modalities:
\begin{align*}
    S_{ij} = S_{ji} = \max(\hat{S}_{ij}, \hat{S}_{ji}), \:
    \hat{S}_{ij} = |\{m: j \in \mathcal{N}_m(i)\}|.
\end{align*}
where $|\cdot|$ is the size of the set.
For instance, if item $j$ is adjacent to item $i$ in both the textual and visual modalities, then $S_{ij} = 2$.
Note that here we symmetrize the similarity matrix, a necessary step to ensure model convergence.

Finally,
the backward stepwise convolution is used to modify the backward optimization process as follows:
\begin{equation}
    \label{eq-bsc}
    \ilu{\emb}_{t} \leftarrow \ilu{\emb}_{t-1} -\eta\cdot \concat_{j=1}^d \sum_{l=0}^L \alpha_{jl}' \mathbf{\tilde{S}}^l [\Delta \ilu{\emb}_t]_{:, j},
\end{equation}
where $\mathbf{\tilde{S}}$ is the similarity matrix after symmetric sqrt normalization and
\begin{equation}
    \label{eq-coef-bsc}
    \alpha_{jl}' = \frac{1 - \beta_j'}{1 - \beta_j'^{L+1}} \beta_j'^l, \:
    \beta_j' = 1 - \beta_j.
\end{equation}
Recall that in the forward pass, STAIR requires the preceding dimensions incorporate more collaborative information while the following dimensions preserve more multimodal information.
The design of $1 - \beta_j$ in Eq.~\eqref{eq-coef-bsc} serves the same purpose, 
ensuring that the later dimensions are more significantly modality-enhanced compared to the preceding dimensions.
Figure~\ref{fig-layer-weights}b illustrates how the layer weights vary across the embedding dimension in comparison to those of the forward pass.

The convergence of Eq.~\eqref{eq-bsc} with fixed $\beta_j \equiv \beta \in [0, 1)$ has been theoretically proved in \cite{xu2024sevo}.
As a straightforward corollary, the convergence rate of BSC after $T$ updates is no worse than $\mathcal{O}\big(1 / ((1 - \max_{j}\beta_j)^2 T) \big)$.

\begin{algorithm}[t]
  \caption{
    STAIR training procedures.
  }
  \label{alg-STAIR}
  \KwIn{
    multimodal features $\mlu{\mathbf{F}}, m \in \mathcal{M}$;
    embedding matrices $\ulu{\emb}, \ilu{\emb}$;
  }
    Initialize $\ilu{\emb}$ by whitening over $\mlu{\mathbf{F}}, m \in \mathcal{M}$;

    Initialize $\ulu{\emb}$ by mean pooling neighboring items;

    Construct kNN similarity graph $\mathbf{S}$ based on Eq.~\eqref{eq_knnk};
    
  \ForEach{step $t$}{
    Compute latent representations $\mathbf{H}$ via FSC;

    Compute BPR loss $\mathcal{L}_{\text{BPR}}$ over $\mathbf{H}$;

    Estimate descent direction at step $t$;

    Commonly update user embeddings;

    Constrainedly  update item embeddings via BSC;
  }
  \KwOut{optimized embeddings $\ulu{\emb}, \ilu{\emb}$.}
\end{algorithm}

\subsection{Training and Prediction}

Given the latent representations $\bm{h}_u, \bm{h}_i \in \mathbb{R}^d$ obtained after the forward stepwise convolution, 
the inner product can be utilized to calculate the scores. 
We adopt the commonly used BPR loss as the training objective; that is,
\begin{equation}
  \mathcal{L}_{\text{BPR}} = -\sum_{u \in \mathcal{U}} \sum_{i \in \mathcal{N}_u} \sum_{i' \not \in \mathcal{N}_u}
  \log \sigma (\bm{h}_u^T \bm{h}_i - \bm{h}_u^T \bm{h}_{i'}),
\end{equation}
where $\mathcal{N}_u$ denotes the interacted items of the user $u$ and $\sigma(\cdot)$ represents the sigmoid function.
The overall training algorithm is detailed in Algorithm~\ref{alg-STAIR}.

\begin{table}
  \centering
  \caption{Dataset statistics}
  \label{table-dataset-statistics}
  \scalebox{0.8}{
  \begin{tabular}{c|cccc}
    \toprule
  Dataset     & \#Users & \#Items & \#Interactions & Sparsity \\
    \midrule
  Baby      & 19,445  & 7,050  & 160,792        & 99.88\%       \\
  Sports        & 35,598  & 18,357  & 296,337        & 99.95\%       \\
  Electronics     & 192,403  & 63,001  & 1,689,188        & 99.99\%       \\
    \bottomrule
  \end{tabular}
  }
\end{table}

\begin{table*}[]
  \centering
  \caption{
    Overall performance comparison.
    The best results are marked in \textbf{bold} and the second-best results are \underline{underlined}.
    Paired t-test is performed over 5 independent runs for evaluating $p$-value ($\le 0.05$ indicates statistical significance).
    Symbol $\blacktriangle$\% stands for the relative improvement against the baselines.
    `-' indicates that the method cannot be performed with a RTX 3090 GPU.
  }
  \label{table-performance}
  \scalebox{.9}{
    \setlength{\tabcolsep}{1.2mm}{
    \begin{tabular}{l|cccc|cccc|cccc}
    \toprule
    & \multicolumn{4}{c|}{Baby} & \multicolumn{4}{c|}{Sports} & \multicolumn{4}{c}{Electronics} \\
     & R@10 & R@20 & N@10 & N@20 & R@10 & R@20 & N@10 & N@20 & R@10 & R@20 & N@10 & N@20 \\
    \midrule
    MF-BPR & 0.0417 & 0.0638 & 0.0227 & 0.0284 & 0.0586 & 0.0885 & 0.0324 & 0.0401 & 0.0372 & 0.0557 & 0.0208 & 0.0256 \\
    LightGCN & 0.0543 & 0.0851 & 0.0293 & 0.0372 & 0.0633 & 0.0956 & 0.0349 & 0.0432 & 0.0393 & 0.0579 & 0.0224 & 0.0272 \\
    JGCF & 0.0563 & 0.0874 & 0.0305 & 0.0385 & 0.0664 & 0.0999 & 0.0369 & 0.0456 & 0.0425 & 0.0618 & 0.0244 & 0.0294 \\ 
    \midrule
    \midrule
    MMGCN & 0.0353 & 0.0580 & 0.0189 & 0.0248 & 0.0331 & 0.0529 & 0.0179 & 0.0230 & 0.0217 & 0.0339 & 0.0117 & 0.0149 \\
    LATTICE & 0.0562 & 0.0875 & 0.0308 & 0.0388 & 0.0683 & 0.1024 & 0.0379 & 0.0466 & - & - & - & - \\
    BM3 & 0.0559 & 0.0866 & 0.0306 & 0.0385 & 0.0647 & 0.0980 & 0.0358 & 0.0444 & 0.0417 & 0.0631 & 0.0233 & 0.0289 \\
    FREEDOM & \underline{0.0649} & \underline{0.0991} & 0.0346 & 0.0434 & \underline{0.0715} & \underline{0.1088} & 0.0383 & 0.0479 & \underline{0.0427} & \underline{0.0647} & \underline{0.0239} & \underline{0.0295} \\
    MMSSL & 0.0595 & 0.0929 & \underline{0.0350} & \underline{0.0442} & 0.0667 & 0.1001 & \underline{0.0390} & \underline{0.0482} & - & - & - & - \\
    \midrule
    \midrule
    STAIR & \textbf{0.0674} & \textbf{0.1042} & \textbf{0.0359} & \textbf{0.0453} & \textbf{0.0743} & \textbf{0.1117} & \textbf{0.0407} & \textbf{0.0503} & \textbf{0.0440} & \textbf{0.0663} & \textbf{0.0245} & \textbf{0.0302} \\
    $\blacktriangle$\% & \multicolumn{1}{c}{3.85\%} & \multicolumn{1}{c}{5.21\%} & \multicolumn{1}{c}{2.55\%} & \multicolumn{1}{c}{2.61\%} & \multicolumn{1}{|c}{3.91\%} & \multicolumn{1}{c}{2.66\%} & \multicolumn{1}{c}{4.36\%} & \multicolumn{1}{c}{4.36\%} & \multicolumn{1}{|c}{3.12\%} & \multicolumn{1}{c}{2.53\%} & \multicolumn{1}{c}{2.60\%} & \multicolumn{1}{c}{2.34\%} \\
    $p$-value & 4.48e-2 & 4.41e-4 & 2.43e-2 & 2.69e-4 & 2.82e-5	& 8.29e-3	& 4.78e-3 & 2.52e-3 & 3.52e-4 & 1.68e-4 & 2.26e-4 & 1.93e-4 \\
    \midrule
    $\blacktriangle$\% over LightGCN & \multicolumn{1}{c}{24.00\%} & \multicolumn{1}{c}{22.44\%} & \multicolumn{1}{c}{22.31\%} & \multicolumn{1}{c}{21.82\%} & \multicolumn{1}{|c}{17.39\%} & \multicolumn{1}{c}{16.83\%} & \multicolumn{1}{c}{16.54\%} & \multicolumn{1}{c}{16.40\%} & \multicolumn{1}{|c}{11.92\%} & \multicolumn{1}{c}{14.56\%} & \multicolumn{1}{c}{9.49\%} & \multicolumn{1}{c}{11.30\%} \\
    $\blacktriangle$\% over FREEDOM & \multicolumn{1}{c}{3.85\%} & \multicolumn{1}{c}{5.21\%} & \multicolumn{1}{c}{3.58\%} & \multicolumn{1}{c}{4.47\%} & \multicolumn{1}{|c}{3.91\%} & \multicolumn{1}{c}{2.66\%} & \multicolumn{1}{c}{6.05\%} & \multicolumn{1}{c}{5.01\%} & \multicolumn{1}{|c}{3.12\%} & \multicolumn{1}{c}{2.53\%} & \multicolumn{1}{c}{2.60\%} & \multicolumn{1}{c}{2.34\%} \\
    \bottomrule
    \end{tabular}
    }
}
\end{table*}

\section{Experiments}

In this section, we present a thorough empirical evaluation 
to validate the effectiveness of our proposed approach and answer the following research questions:
\textbf{RQ1: } Does the proposed STAIR outperform other collaborative filtering methods?
\textbf{RQ2: } Does STAIR enjoy computational and memory efficiency compared with other multimodal models?
\textbf{RQ3: } How do the various modules used in STAIR affect the recommendation performance?
\textbf{RQ4: } How do the key hyperparameters affect the overall performance of STAIR?

\subsection{Experiment Setup}

\paragraph*{Datasets.}
We consider in this paper three commonly used e-commerce datasets obtained from Amazon reviews,
including Baby, Sports, and Electronics.
As suggested by~\cite{zhang2021lattice,zhou2023freedom},
we filter out users and items with less than 5 interactions, 
and Table~\ref{table-dataset-statistics} presents the dataset statistics after preprocessing.
Each dataset contains item thumbnails and text descriptions (\eg, title, brand).
Following~\cite{Zhou2023bm3}, the 4,096-dimensional visual features published in~\cite{ni2019justifying}, and
the 384-dimensional sentence embeddings published in~\cite{zhou2023mmrec} are used for experiments.

\paragraph*{Baselines.}
To validate the effectiveness of STAIR, we compare it with several recommendation models.
Firstly,
MF-BPR~\cite{rendle2012bpr} and two GCN-based methods, LightGCN~\cite{he2020lightgcn} and JGCF~\cite{guo2023jgcf}, 
are chosen as benchmarks in terms of the traditional collaborative filtering that focuses on only the interaction data.
Besides, we consider five multimodal collaborative filtering methods, 
of which LATTICE~\cite{zhang2021lattice} and FREEDOM~\cite{zhou2023freedom} are the two most similar efforts that also use kNN graphs, 
while BM3~\cite{Zhou2023bm3} and MMSSL~\cite{wei2023mmssl} are the two focusing on contrastive learning.

\paragraph*{Evaluation.}
We retrieve the top-$N$ ($N=10, 20$) candidate item list by ranking the scores in descending order. 
Subsequently, two widely used metrics, Recall and Normalized Discounted Cumulative Gain (NDCG), are employed to evaluate the quality of the recommendations. 
The former quantifies the proportion of relevant items within the top-$N$ recommended candidates to all relevant items present in the test set, 
while the latter considers the ranking positions, attributing higher scores to more prioritized items. 
We refer to the average of these metrics across all users as Recall@$N$ (or R@$N$) and NDCG@$N$ (or N@$N$), respectively.

\paragraph*{Implementation Details.}
For fairness~\cite{zhang2021lattice,zhou2023freedom}, 
we fix the embedding dimension to 64 and the number of convolutional layers to 3 for both FSC and BSC processes.
As suggested in \cite{xu2024sevo}, AdamW is employed as the optimizer for training STAIR,
whose learning rate is searched from \{1e-4, 5e-4, 1e-3, 5e-3\} and the weight decay in the range of [0, 1].
The exponent $\gamma$ that controls the changing rate of the layer weights can be chosen from \{0.01, 0.05, 0.1, 0.2, 0.3, 0.4, 0.5\}.
In addition, we adjust the number of neighbors $k_m$ from \{1, 3, 5, 10, 20\} for each modality $m \in \mathcal{M}$ separately.
For the baselines, we determined the optimal learning rate and weight decay in a manner analogous to STAIR, and searched for other hyper-parameters following the suggestions of the original papers.
For all methods,
we report the results on the best checkpoint identified by the validation NDCG@20 metric over 500 epochs.

\subsection{Overall Performance Comparison (RQ1, RQ2)}

\begin{table}[]
  \centering
  \caption{
    Computational and memory costs.
  }
  \label{table-cost}
  \scalebox{0.7}{
  \setlength{\tabcolsep}{1.2mm}{
\begin{tabular}{l|rrr|rrr}
  \toprule
         & \multicolumn{3}{c|}{Time/Epoch (Second)} & \multicolumn{3}{c}{GPU Memory (MB)}     \\
         & Baby   & Sports  & Electronics & Baby   & Sports  & Electronics \\
  \midrule
MF-BPR   & 1.18s  & 1.79s   & 8.33s       & 468M   & 664M    & 1,494M      \\
LightGCN & 1.25s  & 1.99s   & 9.08s       & 478M   & 684M    & 1,676M      \\
JGCF     & 1.31s  & 1.99s   & 11.30s      & 510M   & 742M    & 2,060M      \\
  \midrule
MMGCN    & 2.99s  & 7.84s   & 47.15s      & 1,266M & 2,142M  & 8,530M      \\
LATTICE  & 2.11s  & 11.68s  & -            & 1,664M & 5,928M & -            \\
BM3      & 1.52s  & 3.23s   & 20.81s      & 1,032M & 2,088M  & 6,464M      \\
FREEDOM  & 1.68s  & 3.45s   & 19.41s      & 1,034M & 2,096M  & 6,484M      \\
MMSSL    & 27.70s & 156.80s & -           & 3,048M & 10,656M & -            \\
  \midrule
STAIR    & 1.45s  & 2.65s   & 10.23s      & 490M   & 696M    & 1,738M      \\
  \bottomrule
\end{tabular}
  }
  }
\end{table}

We respectively present the overall performance and efficiency comparison in Table~\ref{table-performance} and \ref{table-cost},
from which we have the following key observations: 

\begin{itemize}[leftmargin=*]

\item 
In traditional collaborative filtering approaches, 
the performance of personalized recommendation continues to improve as the utilization of collaborative information (within interaction data) becomes more and more adequate.
In particular, GCNs play a crucial role in this progress.
However, analogous conclusions cannot be readily inferred from e-commerce multimodal recommendations:
MMGCN performs graph convolution separately on each modality but fails to achieve satisfactory results, 
even inferior to MF-BPR that does not use multimodal features.
This corroborate our argument that how to fully utilize both types of information becomes pivotal in e-commerce multimodal recommendation.

\item
LATTICE and FREEDOM successfully circumvent this dilemma via two separate branches, achieving significantly better results than LightGCN.
However, this approach hinders an optimal trade-off between collaborative and multimodal information, often resulting in too much reliance on one side.
In contrast, STAIR employs a much soft transition to effortlessly fuse both.
The stepwise convolution allows for the co-existence of collaborative and multimodal features along the dimension.
Consequently, STAIR not only surpasses LightGCN by at least 10\%, but also achieves performance improvements ranging from 2\% to 6\% when compared to FREEDOM.
Moreover, as shown in Table~\ref{table-cost}, 
either the learnable similarity matrix of LATTICE or the particular sampling strategy employed by FREEDOM incur non-negligible costs, 
whereas the cost of STAIR is only on par with that of LightGCN.

\item
In contrast to BM3 and MMSSL which emphasize modality learning primarily through contrastive learning, 
the multimodal denoising and enhancement in STAIR are accomplished through a kNN graph and the BSC.
While the former is also capable of mitigating the modality forgetting problem; 
BSC is arguably better adapted to STAIR as it can optimize the embeddings in line with the forward pass.
In addition, we have found that the data augmentation and subsequent computation required by MMSSL can be quite expensive.
When dealing with larger datasets such as Electronics, the computational and memory requirements make MMSSL impossible to implement in real-world recommendations.
In contrast, STAIR has shown to be significantly more efficient.

\end{itemize}

\subsection{Ablation Study (RQ3)}

\begin{figure}[t]
    \centering
    \includegraphics[width=0.4\textwidth]{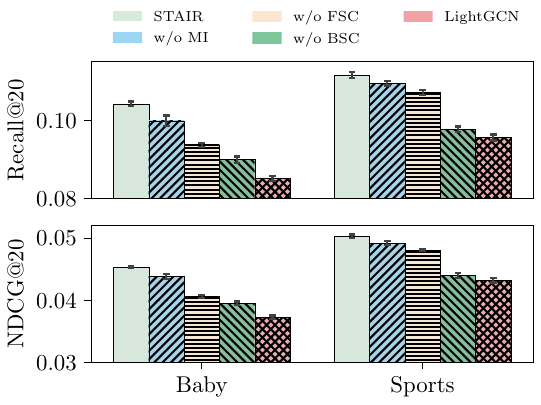}
    \caption{
        STAIR ablation experiments.
    }
  \label{fig-ablation}
\end{figure}


We show the ablation experiments with LightGCN as the baseline in Figure~\ref{fig-ablation}.
First, modality initialization (MI) not only contributes to the final recommended performance, but also improves the robustness to various random seeds.
Second, the usefulness of the forward stepwise convolution (FSC) differs in various datasets.
This can be attributed to the observation that user behavior uncertainty on Sports is lower than that on Baby, leading to a less severe modality erasure issue.
Finally, the backward stepwise convolution (BSC) is essential due to its capability in adaptive modality enhancement to mitigate the forgetting problem.

\subsection{Hyperparameter Sensitivity Analysis (RQ4)}

\begin{figure}[t]
    \centering
    \includegraphics[width=0.43\textwidth]{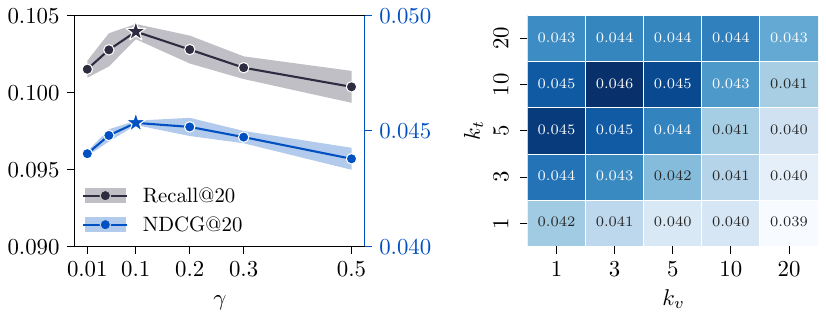}
    \caption{
      Impact of $\gamma$ and $k_m$ on the Baby dataset.
    }
  \label{fig-hyper}
\end{figure}

Figure~\ref{fig-hyper} shows the hyperparameter sensitivity on the Baby dataset.
\textbf{1)} According to Eq.~\eqref{eq_gamma}, 
a larger $\gamma$ implies more emphasis on collaborative information. 
It seems that more multimodal information should be retained during the convolution process, as a slightly lower value of $\gamma$ performs better.
This is because partial collaborative information can be directly learned through the recommendation loss, rather than solely depending on the convolution. 
\textbf{2)} It can be observed that the textual modality is more effective in reflecting users' preferences compared to the visual modality.
This is reasonable since the textual features are encoded based on the less noisy attributes such as title, brand, and category.
This also provides justification for why previous efforts often assign a small weight to the visual modality~\cite{zhou2023freedom}.

\section{Conclusion and Future Work}

We first identify the poor modal-specific user behavior uncertainty in e-commerce scenarios,
and propose modality initialization, forward and backward stepwise graph convolutions for the co-existence of collaborative and multimodal information.
Extensive experiments validate the superiority of STAIR in recommendation accuracy and efficiency.

Note that although STAIR achieves state-of-the-art performance in e-commerce multimodal recommendation,
it may not fully mine the raw multimodal features in content-driven scenarios such as news and video recommendation~\cite{wu2020mind}.
A promising direction in the future is to boost STAIR to benefit from raw multimodal features adaptively.

\section{Acknowledgments}
This work was supported in part by National Natural Science Foundation of China ( No. 62072182 and No. 92270119), Shanghai Institute of Artificial Intelligence for Education, Key Laboratory of Advanced Theory and Application in Statistics and Data Science, Ministry of Education, and East China Normal University.

\bibliography{aaai25}

\newpage
\appendix

\section{Technical Appendix}

\subsection{User Behavior Uncertainty Details}

\begin{figure}[t]
    \centering
    \subfloat[10 clusters]{
        \includegraphics[width=0.46\textwidth]{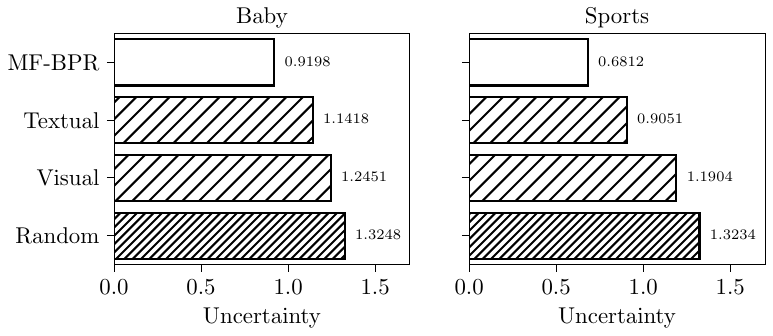}
    }
    \hfil
     \subfloat[20 clusters]{
        \includegraphics[width=0.46\textwidth]{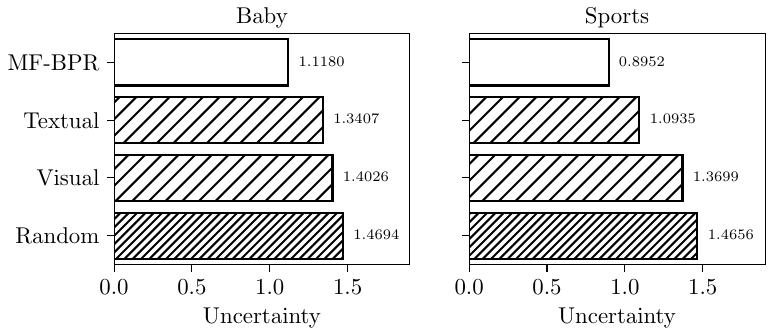}
    }
    \caption{
        User behavior uncertainty under various features. 
        \textbf{MF-BPR}: the embeddings pre-trained via MF-BPR;
        \textbf{Textual/Visual}: the raw multimodal features;
        \textbf{Random}: the randomly generated Gaussian noise.
    }
  \label{fig-uncertainty}
\end{figure}

We detail the steps for estimating user behavior uncertainty based on a specific set of features $\mathbf{F} \in \mathbb{R}^{|\mathcal{I}| \times d'}$.
With the cosine simiarity among the features, K-Means clustering is performed to assign each item an unique class label $c \in 1, 2, \ldots C$, where $C$ denotes the total number of clusters.
Given a user's historical interactions $\mathcal{S}_u = \{i_1, i_2, \ldots, i_{n_u}\}$, the probability of this user favoring the category $c$ can be estimated by
\begin{equation*}
    p_c^{(u)} := \frac{|\{c': c' = c, c' \in \mathcal{S}_u\} |}{n_u}.
\end{equation*}
Subsequently, the behavior uncertainty for this user is defined as the Shannon entropy; that is,
\begin{equation*}
   \mathcal{H}(u) :=  - \sum_{c} p_c^{(u)} \log p_c^{(u)}.
\end{equation*}
As shown in Figure~\ref{fig-uncertainty}, the similar conclusions can be drawn from 10 or 20 clusters.

\subsection{Datasets}

The datasets including Baby, Sports, and Electronics have been preprocessed and published in MMRec~\cite{zhou2023mmrec}.
Particularly, the visual features are encoded via a deep CNN and have been published by~\cite{he2016ups},
while the textual features are encoded from a concatenation of title, descriptions, and categories via a sentence transformer~\cite{reimers2019sentence}.

\subsection{Baselines}

\begin{itemize}[leftmargin=*]
  \item \textbf{MF-BPR}~\cite{rendle2012bpr} operates Matrix Factorization guided by Bayesian personalized ranking (BPR) loss.
  \item \textbf{LightGCN}~\cite{he2020lightgcn} 
  is a pioneering work in traditional collaborative filtering known for its simplicity and effectiveness.
  \item \textbf{JGCF}~\cite{guo2023jgcf} 
  is the state-of-the-art GCN model in collaborative filtering so far,
  with special modifications on manipulating low/high-frequency signals.

  \item \textbf{MMGCN}~\cite{wei2019mmgcn} 
  is a classic multimodal model with modal-specific embeddings and graphs.
  \item \textbf{LATTICE}~\cite{zhang2021lattice} introduces item-item kNN graph based on learnable modality similarity,
  with decoupled collaborative and multimodal convolutions for final representations.
  \item \textbf{BM3}~\cite{Zhou2023bm3} bootstraps latent contrastive views from the representations of users and items, 
  by which it reconstructs the user-item interaction graph and aligns multimodal features.
  \item \textbf{FREEDOM}~\cite{zhou2023freedom} 
  empirically verifies the uselessness of the learnable characteristics of kNN graphs.
  With the a special sampling strategy for denoising, it achieves superior recommendation performance.
  \item \textbf{MMSSL}~\cite{wei2023mmssl} 
  focuses on modality-aware collaborative relation learning and cross-modality dependency modeling to
  enhance the robustness of multimodal recommendation.
\end{itemize}

\subsection{Implementation Details}

\begin{table}[]
    \centering
    \caption{
        Hyperparameters of STAIR over three datasets.
    }
\begin{tabular}{c|ccc}
    \toprule
                    & Baby  & Sports & Electronics \\
    \midrule
$d$ & 64    & 64     & 64          \\
$L$                   & 3     & 3      & 3           \\
Epochs              & 500   & 500    & 500         \\
Batch size          & 1024  & 1024   & 4096        \\
Optimizer           & AdamW & AdamW  & AdamW       \\
Learning rate       & 1e-3  & 1e-3   & 1e-3        \\
Weight decay        & 0.3   & 0.1    & 0.1         \\
$\gamma$               & 0.1   & 0.2    & 0.4         \\
$k_t$                  & 5     & 5      & 5           \\
$k_v$                  & 1     & 1      & 1           \\
    \midrule
\end{tabular}
\end{table}

Apart from the implementation details discussed in the paper.
Here, we provide the settings of STAIR over three datasets.

\end{document}